\documentclass[12pt]{iopart}

\usepackage[style=authoryear-ibid,backend=biber,url=false,doi=true]{biblatex} % Harvardstyle
\usepackage{graphicx, xcolor}
\definecolor{azure}{rgb}{0.0, 0.5, 1.0}

\usepackage{scalerel, tikz}
\usetikzlibrary{calc, svg.path}
\definecolor{orcidlogocol}{HTML}{A6CE39}
\tikzset{
  orcidlogo/.pic={
    \fill[orcidlogocol] svg{M256,128c0,70.7-57.3,128-128,128C57.3,256,0,198.7,0,128C0,57.3,57.3,0,128,0C198.7,0,256,57.3,256,128z};
    \fill[white] svg{M86.3,186.2H70.9V79.1h15.4v48.4V186.2z}
                 svg{M108.9,79.1h41.6c39.6,0,57,28.3,57,53.6c0,27.5-21.5,53.6-56.8,53.6h-41.8V79.1zM124.3,172.4h24.5c34.9,0,42.9-26.5,42.9-39.7c0-21.5-13.7-39.7-43.7-39.7h-23.7V172.4z}
                 svg{M88.7,56.8c0,5.5-4.5,10.1-10.1,10.1c-5.6,0-10.1-4.6-10.1-10.1c0-5.6,4.5-10.1,10.1-10.1C84.2,46.7,88.7,51.3,88.7,56.8z};
  }
}
\newcommand\orcidlink[1]{\href{https://orcid.org/#1}{{\mbox{\scalerel*{
    \begin{tikzpicture}[yscale=-1,transform shape]
    \pic[]{orcidlogo};
    \end{tikzpicture}
}{-}}}}} %this symbol controls the size

\usepackage{url}
\usepackage[colorlinks=true, 
            citecolor=azure, 
            linkcolor=azure, %figures, anchors etc
            urlcolor=azure]{hyperref} %autoref

\usepackage[nolist,nohyperlinks]{acronym}

\begin{acronym}[ECU]

\acro{AWMPS}[AWMPS]{arbitrary waveform magnetic particle spectrometer}
\acro{ADC}[ADC]{analog-to-digital converter}
\acro{AUC}[AUC]{area under the curve}

\acro{CT}[CT]{computed tomography}

\acro{DAC}[DAC]{digital-to-analog converter}
\acro{DFG}[DFG]{drive field generator}

\acro{ECD}[ECD]{equivalent circuit diagram}

\acro{FFL}[FFL]{field-free-line}
\acro{FFP}[FFP]{field-free-point}
\acro{FFR}[FFR]{field-free-region}
\acro{FFT}[FFT]{fast Fourier transform}
\acro{FOV}[FOV]{field of view}
\acro{FWHM}[FWHM]{full width at half maximum}

\acro{GUI}[GUI]{graphic user interface}

\acro{HCC}[HCC]{high current circuit}

\acro{ICN}[ICN]{inductive coupling network}
\acro{ISI}[ISI]{integrated signal intensity}
\acro{ICs}[ICs]{integrated circuits}

\acro{LFR}[LFR]{low-field-region}
\acro{LNA}[LNA]{low noise amplifier}

\acro{MPI}[MPI]{Magnetic Particle Imaging}
\acro{MRI}[MRI]{Magnetic Resonance Imaging}
\acro{MTT}[MTT]{mean transit time}

\acro{PSF}[PSF]{point spread function}
\acro{PNS}[PNS]{peripheral nerve stimulation}
\acro{PTT}[PTT]{pulmonary transit time}

\acro{Q}[Q-factor]{quality factor}

\acro{RF}[RF]{radio-frequency}
\acro{RF-fields}[RF]{radio frequency fields}
\acro{rBV}[rBV]{relative blood volume}
\acro{rBF}[rBF]{relative blood flow}
\acro{RBCs}[RBCs]{red blood cells}

\acro{SNR}[SNR]{signal-to-noise ratio}
\acro{SAR}[SAR]{specific absorption rate}
\acro{SPIONs}[SPIONs]{superparamagnetic iron oxide nanoparticles}

\acro{TF}[TF]{transfer function}
\acro{THD}[THD]{total harmonic distortion}
\acro{TTP}[TTP]{time to peak}

\acro{VOI}[VOI]{volume of interest}

\end{acronym}

\usepackage[exponent-product = \cdot]{siunitx}
\renewcommand{\mat}[1]{\boldsymbol{#1}}
\newcommand{\IR}[0]{\mathbb{R}}

\usepackage{iopams}  

\begin{document}

\title{Saline bolus for negative contrast perfusion imaging in magnetic particle imaging}

\author{Fabian Mohn\textsuperscript{1,2}\orcidlink{0000-0002-9151-9929},
Miriam Exner\textsuperscript{2}\orcidlink{0000-0002-3756-3822},
Patryk Szwargulski\textsuperscript{2}\orcidlink{0000-0003-2563-9006},
Martin M\"oddel\textsuperscript{1,2}\orcidlink{0000-0002-4737-7863},
Tobias Knopp\textsuperscript{1,2}\orcidlink{0000-0002-1589-8517},
Matthias Graeser\textsuperscript{1,2,3,4}\orcidlink{0000-0003-1472-5988}
}

\address{\textsuperscript{1} Section for Biomedical Imaging , University Medical Center Hamburg-Eppendorf, Hamburg, Germany}
\address{\textsuperscript{2} Institute for Biomedical Imaging, Hamburg University of Technology, Hamburg, Germany}
\address{\textsuperscript{3} Fraunhofer Research Institution for Individualized and Cell-based Medicine, IMTE, L\"ubeck, Germany}
\address{\textsuperscript{4} Institute for Medical Engineering, University of L\"ubeck, L\"ubeck, Germany}

\ead{\href{mailto:fabian.mohn@tuhh.de}{fabian.mohn@tuhh.de}}
\vspace{10pt}
\begin{indented}
\item[]June 2023, R1
\end{indented}

%%%%%%%%%%%%%%%%%%%%%%%%%%%%%%%%%%%%%%%%%%%%%%%%%%%%%%%%%%%%%%%%%%%%%%%%%
\begin{abstract} 
\textit{Objective.}
\ac{MPI} is capable of high temporal resolution measurements of the spatial distribution of magnetic nanoparticles and therefore well suited for perfusion imaging, which is an important tool in medical diagnosis.
Perfusion imaging in \ac{MPI} usually requires a fresh bolus of tracer material to capture the key signal dynamics.
Here, we propose a method to decouple the imaging sequence from the
injection of additional tracer material, without further increasing the administered iron dose in the body with each image. 
\textit{Approach.}
A bolus of physiological saline solution without any particles (negative contrast) diminishes the steady-state concentration of a long-circulating tracer during passage. This depression in the measured concentration contributes to the required contrast dynamics. The presence of a long-circulating tracer is therefore a prerequisite to obtain the negative contrast.
As a quantitative tracer based imaging method, the signal is linear in the tracer concentration for any location that contains nanoparticles and zero in the surrounding tissue which does not provide any intrinsic signal. 
After tracer injection, the concentration over time (positive contrast) can be utilized to calculate dynamic diagnostic parameters like perfusion parameters in vessels and organs. Every acquired perfusion image thus requires a new bolus of tracer with a sufficiently large iron dose to be visible above the background.
\textit{Main results.}
Perfusion parameters are calculated based on the time response of the proposed negative bolus and compared to a positive bolus. Results from phantom experiments show that normalized signals from positive and negative boli are concurrent and deviations of calculated perfusion maps are low.
\textit{Significance.}
Our method opens up the possibility to increase the total monitoring time of a future patient by utilizing a positive-negative contrast sequence, while minimizing the iron dose per acquired image. 
\end{abstract} 
%%%%%%%%%%%%%%%%%%%%%%%%%%%%%%%%%%%%%%%%%%%%%%%%%%%%%%%%%%%%%%%%%%%%%%%%%

\vspace{2pc}
\noindent{\it Keywords}: Magnetic Particle Imaging, perfusion imaging, magnetic tracer, negative contrast, long term monitoring \\[2pc]

%%%%%%%%%%%%%%%%%%%%%%%%%%%%%%%%%%%%%%%%%%%%%%%%%%%%%%%%%%%%%%%%%%%%%%%%%
%%%%%%%%%%%%%%%%%%%%%%%%%%%%%%%%%%%%%%%%%%%%%%%%%%%%%%%%%%%%%%%%%%%%%%%%%
\section{Introduction} 
\acresetall

\ac{MPI} is a tracer based imaging modality which measures the spatial concentration of magnetic nanoparticles \autocite{gleich_tomographic_2005}. After rapid development, the method has already proven its potential in several clinical applications like control of the temperature rise in magnetic hyperthermia~\autocite{murase_control_2013}, monitoring of cellular grafts~\autocite{zheng_quantitative_2016}, cancer detection~\autocite{yu_magnetic_2017-1}, quantification of vascular stenosis~\autocite{vaalma_magnetic_2017}, 
lung perfusion imaging~\autocite{zhou_first_2017}, traumatic brain injury imaging~\autocite{orendorff_first_2017} and stroke detection~\autocite{ludewig_magnetic_2017,szwargulski_monitoring_2020}. Recent developments show that MPI is feasible on a human scale~\autocite{rahmer_remote_2018,graeser_human-sized_2019,mason_concept_2021,vogel_impi_2022,thieben_safe_2023}.

\ac{MPI} uses the nonlinear magnetization behaviour, which is controlled by the superposition of a static spatial encoding field (selection field) and an oscillating homogeneous field (drive field). The first creates a \ac{FFR}, the second deflects the \ac{FFR} and causes the particles to realign to the current field vector, which can be measured using Faraday's law of induction \autocite{gleich_tomographic_2005}. 
Due to the non-linearity of the particles, the receive signal contains higher harmonics, encoding a location depend spectral fingerprint. Using Fourier transform, a solution to the underlying system of linear equations reveals the location and concentration of the tracer \autocite{gleich_tomographic_2005}.

As MPI depends on a magnetic tracer injected into the human body, several questions regarding the safety of these tracers arise. The iron oxide particles are usually coated either by dextran or PEG shells making the tracer compatible to the body~\autocite{reimer_ferucarbotran_2003,arbab_model_2005,oh_iron_2011}. If coated with dextran, the particles are taken up by Kupffer-cells in short time intervals ($\approx$ \SI{10}{\min} half-life) \autocite{reimer_ferucarbotran_2003,haegele_magnetic_2014} and are integrated in the human iron pool by the liver. Certain PEG shells remain longer within the blood pool with a circulation half-life of approximately \SI{7}{\hour} in mice~\autocite{liu_long_2021}. After that, these particles are also taken up by the liver and spleen~\autocite{liu_long_2021}. A long half-life is as well achieved by labeling \acl{RBCs} with magnetic particles, where 30\% of signal remains after \SI{24}{\hour} \autocite{antonelli_red_2013}. 
Once in the liver, the particles take up to several weeks to be dismantled \autocite{reimer_ferucarbotran_2003}. Therefore, in the case of MPI, the maximum iron dose is typically a threshold for a safe use of magnetic tracer in humans. The range of iron doses for human use varies from 2.24\,mg$_\textup{Fe}$/kg (Resovist) to 8.5\,mg$_\textup{Fe}$/kg (Feraheme)~\autocite{southern_commentary_2018}. As both are designed and medically approved for very different purposes, a specific MPI tracer has yet to be approved for imaging in humans. Nevertheless, the doses of these tracers can be used as an indication for safety in humans. 
Concerns remain if particles can be fully incorporated naturally or if toxicity remains in long-term metabolism~\autocite{billings_magnetic_2021,rubia-rodriguez_whither_2021,sun_magnetic_2008}, which motivates to reduce the added iron amount as much as possible.

In \textcite{graeser_human-sized_2019}, a surveillance scenario for a stroke patient was presented by investigating brain perfusion imaging. Perfusion imaging is an important technique to provide hemodynamics which are used to characterize pathological conditions~\autocite{cha_perfusion_2003,wintermark_comparative_2005}.
For perfusion imaging, usually short circulating tracers are used, because long circulating tracers would only lead to a higher background signal after distribution within the body. In such a scenario, a bolus injection is given in every imaging sequence, e.g. when monitoring a stroke patient. To comply with the above-mentioned limits, each single bolus may only contain a fraction of the total iron dose, with a minimum dose based on the system's sensitivity. Also, each bolus needs to be visible above the background of prior boli (baseline concentration). Thus, each additional bolus reduces the number of future images that can be acquired safely. 

In order to calculate perfusion parameters, the shape, peak, timing and \ac{AUC} of an increase in concentration is assessed~ 
\autocite{ludewig_magnetic_2017,graeser_human-sized_2019,graeser_design_2020,kaul_pulmonary_2021,ludewig_magnetic_2022,ostergaard_high_1996,ostergaard_principles_2005,cha_perfusion_2003,wintermark_comparative_2005}.
Other work focused on blood flow velocity \autocite{kaul_magnetic_2018}, kidney perfusion \autocite{molwitz_first_2019} or blood flow with stenosis \autocite{siepmann_image-derived_2021}, using similar methods.
Apart from an increase in concentration, perfusion parameters may also be obtained from a concentration decrease, known from \ac{MRI} as negative contrast~\autocite{detre_perfusion_1992,barbier_methodology_2001}. Other works in \ac{MPI} also took advantage of a decrease in concentration to identify and track changes e.g. for the inflation condition of a balloon catheter~\autocite{salamon_magnetic_2016}.   

In this work, we present an approach that will enable long-term monitoring of a future patient, without increasing the administered iron dose in the body with every image.
Using long circulating tracers, we decouple the imaging sequence from the injection of additional tracer material. To acquire the necessary dynamics in the signal, from which perfusion parameters are calculated, we use physiological neutral saline solution to shortly reduce the concentration of the tracer circulating within the perfused tissue (negative contrast). After image processing, this depression in concentration can be used in the same way as the positive change in concentration of a bolus to derive perfusion parameters. The underlying principle requires a minimum baseline or steady-state concentration of tracer to be present in the \ac{VOI}.

A circulatory experimental setup is employed in our study, with a rat-scaled heart phantom on a pre-clinical \ac{MPI} system.
We show that signals from negative and positive boli are concurrent, that the deviations in shape and \ac{AUC} are low after normalization and that perfusion maps can be obtained from negative boli. 
This method has great potential to increase the number of diagnostic perfusion examinations of a future patient within the safety limits of the total iron dose.

This paper is the full and extended version of our conference abstract \cite{mohn_using_2023}.

%%%%%%%%%%%%%%%%%%%%%%%%%%%%%%%%%%%%%%%%%%%%%%%%%%%%%%%%%%%%%%%%%%%%%%%%%
%%%%%%%%%%%%%%%%%%%%%%%%%%%%%%%%%%%%%%%%%%%%%%%%%%%%%%%%%%%%%%%%%%%%%%%%%
\section{Methods}

We consider a scenario that  requires a baseline tracer concentration, which could be reached with either a single large initial bolus, by labeled and administered \acl{RBCs} or accumulated over time by several smaller positive boli. Once the tracer concentration has reached a homogeneous steady-state in the circulating blood pool, a bolus of saline solution can be imaged due to the negative contrast (a reduced signal over time) that is created by tracer dilution. To prove this hypothesis, perfusion parameters are derived from the time response of negative and positive boli. Positive boli serve as the ground truth in this work and verification is done by comparing dynamic subtraction images, normalized time curves and perfusion maps between negative and positive bolus types, namely the \ac{TTP}, \ac{rBF}, \ac{rBV} and \ac{MTT}.
In this work, we administer an initial large positive bolus for a baseline concentration and continue with alternating small positive and negative boli, and measure the time depended concentration variations using \ac{MPI}.

%%%%%%%%%%%%%%%%%%%%%%%%%%%%%%%%%%%%%%%%%%%%%%%%%%%%%%%%%%%%%%%%%%%%%%%%%
\subsection{Experimental setup}

\begin{figure}[t!]
    \centering
    \includegraphics[width=1.0\linewidth]{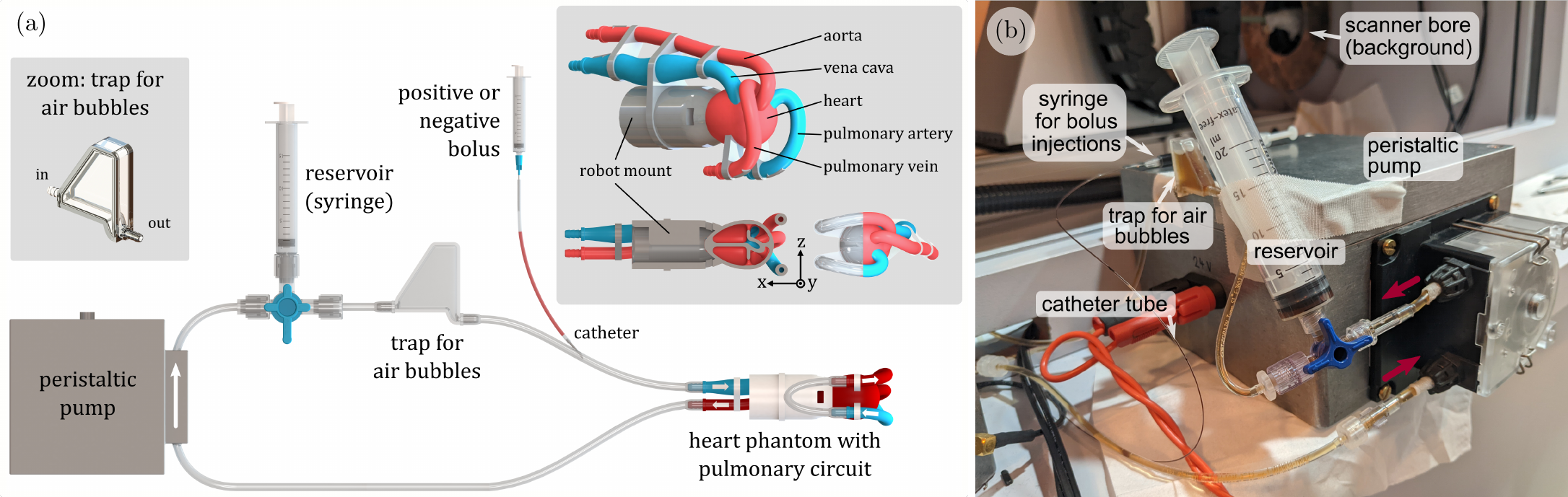}
    \caption{\textbf{Experimental Setup.} A visualization of the experimental setup in (a), stated in flow direction: peristaltic pump, reservoir (syringe), air bubble trap, bolus injection point, 3D printed heart phantom, return tube. Only the heart phantom is located inside the scanner bore. The bubble trap and the phantom are shown as a blow-up in the upper corners. 
    To the right (b), a labeled photo of the setup at the scanner is shown. The tubes reach inside the scanner bore in the background where the phantom is placed. All other parts from (a) are visible.}
    \label{fig:expSetup}
\end{figure}

To validate the feasibility of negative boli, we constructed a rat-sized heart phantom for circulatory flow experiments, shown in \autoref{fig:expSetup}.
The hollow phantom is connected to a peristaltic pump using an average flow rate of \SI{31}{\ml\min^{-1}}, a realistic cardiac output for a rat~\autocite{treuting_comparative_2018}, to circulate the imitated blood pool (desalinated water). 
The heart phantom has a volume of \SI{1763}{mm^3}, with a portion of 60\% for the actual heart consisting out of 4 chambers (left/right atrium/ventricle). All partial volumes are realistically scaled and designed on the basis of a \SI{200}{\g} rat~\autocite{exner_3d_2019}. Different details of the phantom are displayed in \autoref{fig:expSetup}\,(a). The total volume of the imitated blood pool is \SI{14}{\ml} including all tube lengths and reservoirs, similar to the blood volume of a 200 g rat~\autocite{belcher_studies_1957}. The phantom was placed in the center of the MPI system via a robotic mount.
Following the experimental setup shown in \autoref{fig:expSetup}, the fluid passes a reservoir of \SI{1}{\ml} first, to compensate pressure and changes in volume when a bolus is injected. Next is a 3D printed trap for air bubbles, which filters arising air from pressure variation or by injection to keep the phantom free of trapped air. The injection point is located at a realistic distance (vena caudalis) from the heart phantom, where a small catheter tube (\SI{0.28}{\mm} diameter) is connected to the bolus syringe.
The pulmonary circuit consists of a loop with a tube length of \SI{9}{\cm} and \SI{1.6}{mm} diameter to achieve a delay of \SI{1}{s}, which is the \ac{PTT} of a healthy rat \autocite{su_comparison_2022}. It should be noted that for simplicity in these flow experiments, tissue mimicry to simulate real organ perfusion was neglected at this small organ scale.
Before measurements, the peristaltic pump was fitted in a shielded housing and measurements were performed to exclude any influence of the motor on the MPI measurement signal.

%%%%%%%%%%%%%%%%%%%%%%%%%%%%%%%%%%%%%%%%%%%%%%%%%%%%%%%%%%%%%%%%%%%%%%%%%
\subsection{Phantom measurements}
\label{sec:phanmeas}

The measurements were performed using a Bruker preclinical MPI system (MPI FF 20/25, Bruker BioSpin MRI GmbH, Ettlingen Germany) with a gradient field set to \SI{1.2}{\tesla\per\meter} in $z$-direction and a drive-field amplitude of \SI{12}{\milli\tesla} in all spatial directions. Consequently, the image \ac{FOV} spans \qtyproduct{40x40x20}{\mm}. For detection, a 3D gradiometric receive coil with an open bore of \SI{72}{\mm} was used, which enables high \ac{SNR} and sensitivity~\autocite{paysen_improved_2018}, similar to the one constructed in \autocite{graeser_towards_2017}. Measurements of $12000$ frames were recorded without averaging at \SI{21.54}{\ms} repetition time to observe changes in the concentration over \SI{4.3}{\min} per bolus.
A baseline concentration of \SI{237}{\ug_{Fe}}\si{\per\ml} (\SI{4.24}{\milli\mol\per\l}) Perimag (micromod GmbH, Rostock, Germany) was chosen for the \SI{14}{\milli\l} imitated blood pool volume and administered via a single initial bolus which circulated until homogeneously dispersed (in total \SI{3.3}{\mg} iron).
During measurements, positive and negative boli were applied alternatingly to minimize concentration changes between boli. A total of $8$ boli was given, $4$ boli of each type. The time between boli was set to approximately \SI{5}{\min} while the pump continuously circulated the pool to reach a homogeneous distribution before the next injection.
A volume of \SI{150}{\uL} was chosen for all boli, either as a tracer dispersion of \SI{1}{\ug_{Fe}}\si{\per\ul} (positive) or consisting of physiological neutral saline solution (negative). Before injection of another bolus, \SI{150}{\uL} were drawn out of the pool to guarantee identical starting conditions.
The injection was queued in the catheter tube, with a leading tiny air pocket to avoid dispersion of the bolus during preparation and to obtain identical constant injection rates.
By means of the long catheter tube, the bolus can be conveniently applied from outside the scanner. A syringe-pump was not used due to the high pressure necessary to operate the small bolus syringe.
Ongoing experiments were supervised with an online reconstruction software~\autocite{knopp_online_2016} to evaluate success (bolus passages) and confirm a homogeneous steady-state distribution.

%%%%%%%%%%%%%%%%%%%%%%%%%%%%%%%%%%%%%%%%%%%%%%%%%%%%%%%%%%%%%%%%%%%%%%%%%
\subsection{Image reconstruction and post processing}

All images were reconstructed using the system matrix approach described in \textcite{gleich_tomographic_2005}. The system matrix was recorded on a \qtyproduct{22x22x22}{} grid, spanning a system matrix \ac{FOV} of \qtyproduct{44x44x22}{\mm}. This includes an overscan of 1 voxel compared to the image \ac{FOV} and yields a voxelsize of \qtyproduct{2x2x1}{\mm}. Each single point response  was averaged 150 times.
Solutions were obtained by solving a Tikhonov regularized least squares problem using the MPI reconstruction framework MPIReco.jl~\autocite{knopp_mpirecojl_2019}. For reconstruction, a relative regularization parameter of $\lambda = 0.01$ and $5$ iterations were chosen. Frequency selection includes frequencies from \SI{80}{\kHz} to \SI{1.25}{\MHz} with an \ac{SNR} of at least $2$. A background measurement (without the phantom in the scanner) was subtracted from the phantom measurements before reconstruction.  
The reconstructed solution $\mat{c} \in \IR^{N\times n_t}$ with $N=n_x \times n_y \times n_z =22^3$ voxels has $n_t=12000$ time frames.  
The term $c_{i,t}$ refers to the $i$-th voxel and $t$-th time frame of the discrete solution.
Furthermore, $\mat{c}_t^{\textup{bolus}\,j}$ refers to all voxels of the $j$-th bolus.

\begin{figure}[t!]
    \centering
    \includegraphics[width=0.9\linewidth]{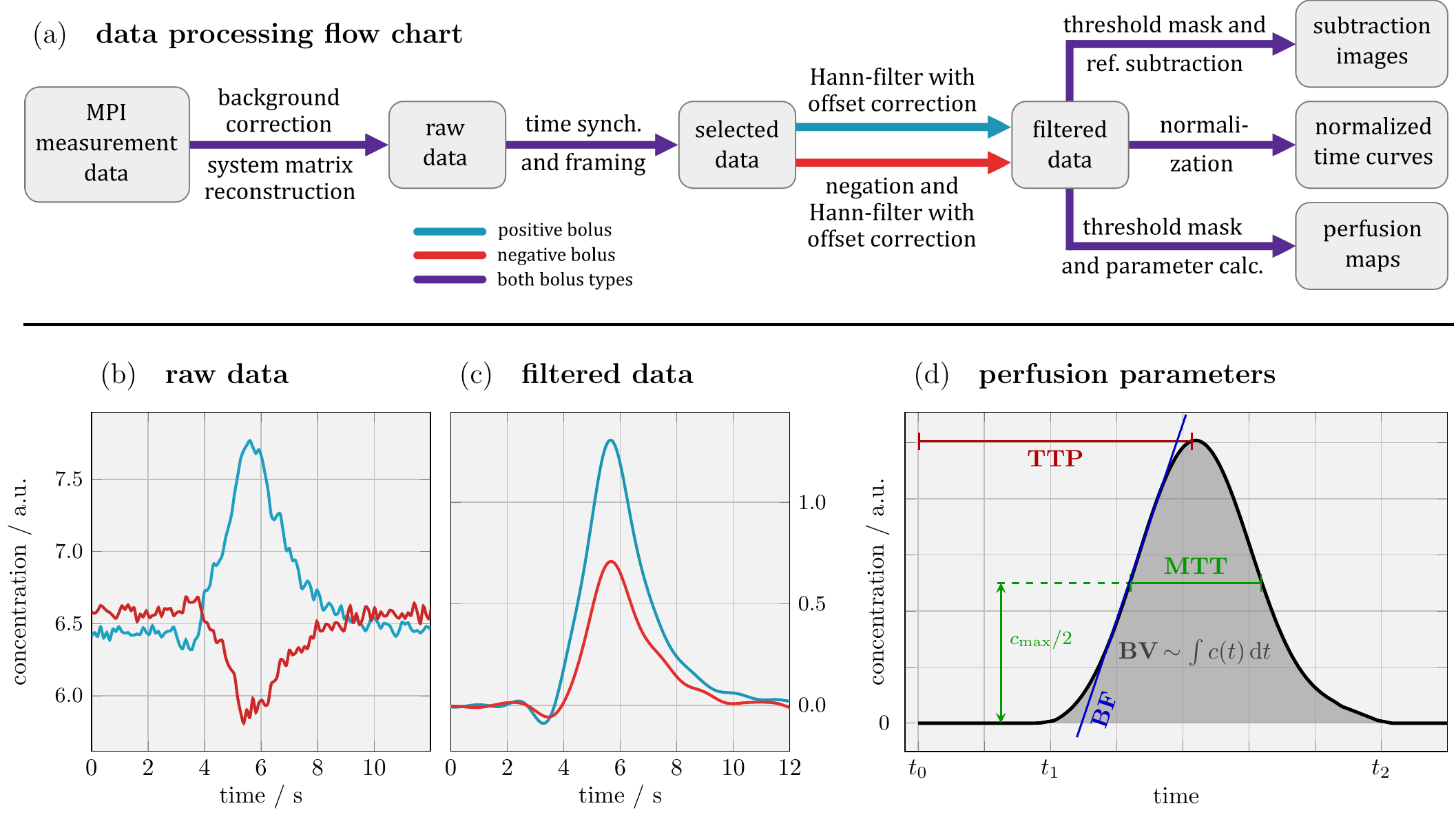} 
    \caption{\textbf{Processing flow chart and signal analysis.} The data processing flow chart is shown in (a). The measurement data is reconstructed, the relevant time frame containing the first passing of the bolus (12\,s) is selected, then negated (negative bolus only), and filtered before post processing. A threshold mask selects only relevant voxels. Raw data and filtered data are shown exemplarily for one voxel in (b) and (c), respectively. In (d), the applied definitions of perfusion parameters are visualized.}
    \label{fig:methods}
\end{figure}

The entire data processing chain is displayed as a flow-chart in \autoref{fig:methods}, along with data examples and parameter definitions.
After reconstruction, the post processing includes time synchronization, data negation, Hann-filtering, voxel masking and lastly the perfusion parameter calculation. 
In the order of mention: the time is synchronized over all boli by selecting a time frame that starts with the injection ($t_0$), includes the first bolus ($t_1$ to $t_2$) and stops after the first passing ($t=12$\,s). This step ensures that only relevant data is processed later.
For negative boli the concentration is inverted, so they can be processed identically to positive boli by the same code that works e.g. by using the maximum intensity. 
Afterwards, an appropriate filter type was chosen to smoothen the data for a more accurate peak detection and data interpolation, which also shifts the concentration offset to zero. To reject ringing artifacts, we avoided rectangular windows and selected a low-pass Hann-filter with a window-size of $10$ samples. The Hann-filter is applied voxel-wise on the Fourier transformed temporal data.
The last step before perfusion parameters are calculated is the determination of a threshold mask, which reduces image noise by excluding any voxel with an intensity lower than $6$\% of the maximum value for the selected time frame.
The term "raw data" refers to data reconstructed from the measurement signal before any processing, and "filtered data" is a term for the final Hann-filtered data prior to post processing. This last step includes subtraction images, normalization and perfusion maps.

\paragraph{Subtraction images.}
To visualize small changes in concentration, a pre-bolus reference is subtracted (pre contrast) from all following images (post contrast) to remove the native concentration, and consequently only changes in the concentration become visible~\autocite{hecht_renal_2004}.
Dynamic subtraction images are based on the filtered data and calculated by subtracting with the identical spatial slice at $t_0$, that is before the administered bolus reaches the \ac{FOV}.

\paragraph{Normalized images.} 
By using a normalization, the time response of positive and negative boli can be overlaid e.g. to compare rise times or the shape of the bolus.
To this end, the filtered data was used and normalization was applied to match maxima (in this graph only).
The normalization is done via linear regression to match the maxima of the first peak without distorting the underlying time axis.
Based on the linear model $\mat{c}_t^{\textup{bolus}\,1} = a_j \mat{c}_t^{\textup{bolus}\,j}+b_j$ for all $j \in \{2,\dots,8\}$ boli, a slope $a_j$ and an offset correction $b_j$ is determined for all time samples $t$, so each bolus can be mapped individually to one selected positive bolus~\autocite{zou_correlation_2003}.

\paragraph{Perfusion images.}
In order to assess the feasibility of negative boli in perfusion imaging, calculations were performed on the filtered data for positive and negative boli using a threshold mask as described above and in \autoref{fig:methods}\,(d).
The data is not normalized for this calculation and the identical functions were used for positive and negative boli. The only difference is that negative boli are inverted before filtering.
Perfusion maps are generated for \ac{TTP}, \ac{rBF}, \ac{rBV} and \ac{MTT}. The definition of individual perfusion parameters follows in \autoref{sec:perfparams} below.

%%%%%%%%%%%%%%%%%%%%%%%%%%%%%%%%%%%%%%%%%%%%%%%%%%%%%%%%%%%%%%%%%%%%%%%%%
\subsection{Perfusion parameters}
\label{sec:perfparams}

The following definitions are formulated continuously by using $c_i(t)$, to provide a more general consideration. However, our data were processed using the discrete solution $c_{i,t}$.

\paragraph{TTP:} 
The \acf{TTP} is defined as the time elapsed between a chosen reference point (the bolus injection, here $t_0$) and the measured signal maximum of the first bolus passing (see \autoref{fig:methods}\,(d)). The \textbf{TTP} $\in \IR^N$ is calculated element-wise for all voxel $i \in \{1,\dots,N\}$ via
\begin{equation}
    \textup{TTP}_i = \textup{arg\,max}_t \left(\, c_i (t)\, \right),
\end{equation}
where $c_{i}(t)$ is the concentration over time of the $i$-th voxel~\autocite{fieselmann_deconvolution-based_2011}. 

\paragraph{PTT:} 
\noindent The \acf{PTT} is the time that a particle or blood cell needs to pass from the right to the left ventricle, therefore the average time that the blood circulates through the lung. It can be calculated by the difference of the \ac{TTP}s of the ventricles~\autocite{dean-ben_high-frame_2015}. In this work it is defined by a length of tube and the flow rate that was chosen to introduce a delay of about $1$\,s (see \autoref{sec:phanmeas}).

\paragraph{rBF:} 
\noindent The \acf{rBF} equals the highest positive gradient in $c_{i}(t)$ 
\begin{equation}
    \textup{rBF}_i = \textup{arg\,max}_t \left( \, \frac{d}{dt} c_i(t) \, \right ),
\end{equation}
as shown in \autoref{fig:methods}\,(d). A calculation for all voxels yields \textbf{rBF} $\in \IR^N$.
The BF and BV are calculated in a relative manner, due to a missing correct arterial input function.

\paragraph{rBV:} 
\noindent The \acf{rBV} can be derived from an element-wise evaluation of the \ac{AUC} in the $i$-th voxel, divided by the \ac{AUC} in the artery to yield \textbf{rBV} $\in \IR^N$. Therefore, the blood volume is proportional to the integral of $c_i(t)$, divided by the integral of $c_\textup{art}(t)$ in the artery, over their respective time intervals $[t_1,t_2]$ (see \autoref{fig:methods}\,(d)), that mark the time the bolus needs for the first passing, as in
\begin{equation}
    \textup{rBV}_i = \frac{\int_{t_1}^{t_2} c_i(t) dt} {\int_{t_1}^{t_2} c_{\textup{art}}(t) d t} ~.
\end{equation}

\paragraph{MTT:} 
\noindent The \acf{MTT} is a measure of the average time that a particle or blood cell spends inside a vessel or organ. It is usually defined by the ratio of \ac{rBV} to \ac{rBF} or via the first moment of the \ac{AUC} \autocite{weisskoff_pitfalls_1993,ostergaard_high_1996}, however, it strongly correlates with the \ac{FWHM} of the concentration peak~\autocite{kealey_user-defined_2004,ostergaard_principles_2005}, especially for measurements with low tissue perfusion. To avoid error propagation due to inaccuracies in \ac{rBV} and \ac{rBF}~\autocite{weisskoff_pitfalls_1993,ostergaard_high_1996}, we took the time interval \ac{FWHM} as the \ac{MTT}~\autocite{ostergaard_principles_2005}, shown in \autoref{fig:methods}\,(d).

\subsection{Comparison of positive and negative boli performance}
Negative and positive boli are qualitatively compared by visual inspection of subtraction images, normalized data and perfusion maps. 
All these calculations are based on the filtered data set, but with an increasing computational effort in the mentioned order. The subtraction images reveal small changes in the concentration over time and require just the filtered time data and a threshold mask. Normalized images require a normalization step via linear regression and perfusion maps need to be processed according to \autoref{sec:perfparams}. All of these calculations were done for all boli, however, in the results only some boli and some perfusion maps are shown exemplarily. Positive boli are taken as the ground truth during comparison to negative boli in this work.

\begin{figure}[t!]
    \centering
    \includegraphics[width=0.85\linewidth]{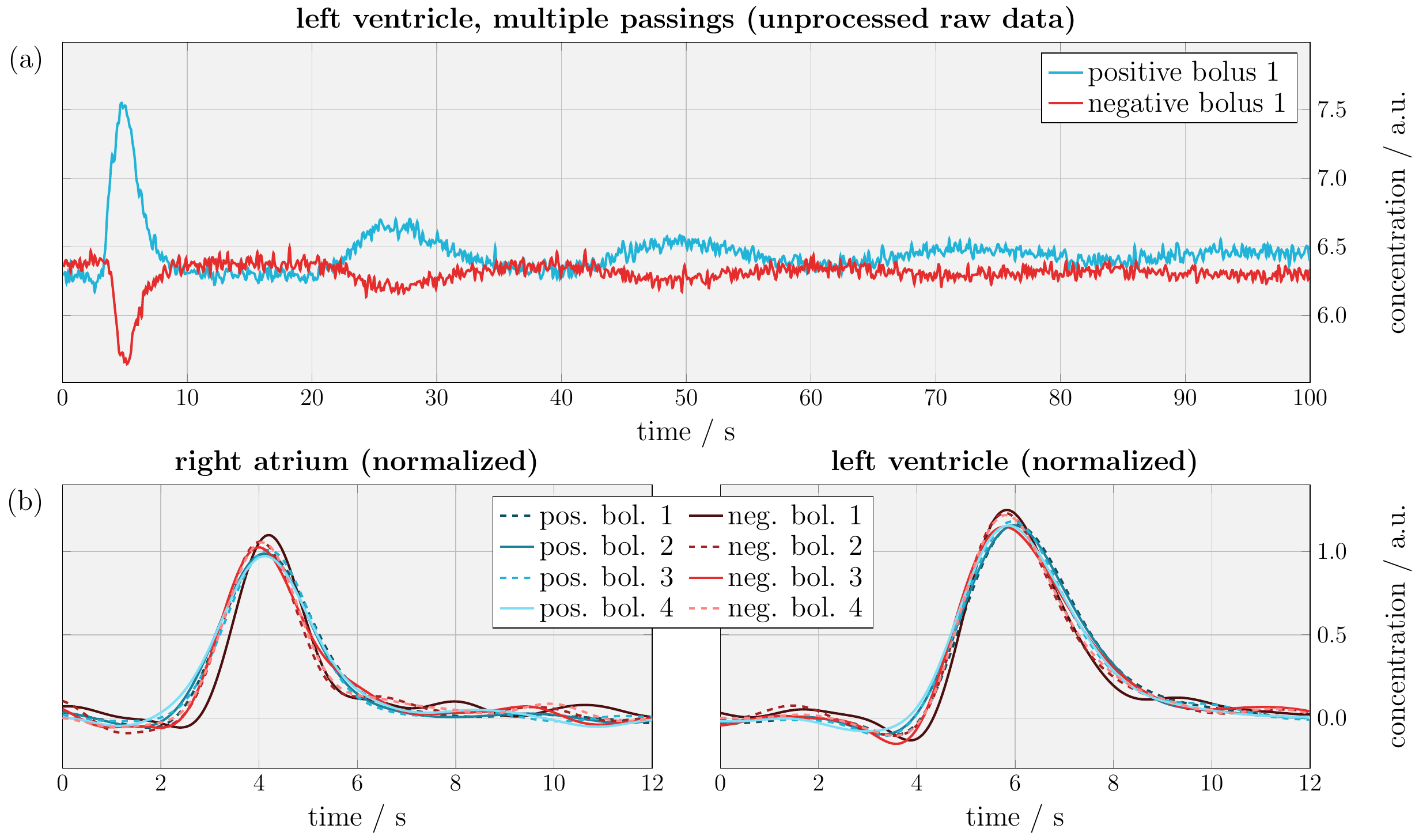} 
    \caption{\textbf{Raw data and normalized data.} In (a), the exemplary raw data of a consecutive positive and negative bolus for $100$\,s are shown. Several bolus passages are visible, indicated by numerous peaks. The baseline steady-state concentration of the negative bolus (prior to first pass) is increased due to the preceding positive bolus.
    Below in (b), the right atrium and the left ventricle of the filtered data including the normalization step are shown for all $8$ boli.}
    \label{fig:timeResp_raw}
\end{figure}

%%%%%%%%%%%%%%%%%%%%%%%%%%%%%%%%%%%%%%%%%%%%%%%%%%%%%%%%%%%%%%%%%%%%%%%%%
%%%%%%%%%%%%%%%%%%%%%%%%%%%%%%%%%%%%%%%%%%%%%%%%%%%%%%%%%%%%%%%%%%%%%%%%%
\section{Results}

\subsection{Comparison of signal time histories, difference images, and normalization of the positive and negative boli}
\label{sec:res_time}

In \autoref{fig:timeResp_raw}\,(a), raw data of positive and negative boli in the left ventricle are shown, displaying several bolus passages over \SI{100}{\s}. The steady-state level increases slightly after each additional positive bolus (boli are administered alternately).
Due to the relation of the chosen bolus concentration and initial baseline, peaks of negative boli are about $43$\% lower in the raw data. 

Below in \autoref{fig:timeResp_raw}\,(b), normalized curves based on filtered data of all 8 boli are shown for the right atrium and left ventricle. The normalized overlay of all 8 boli at the same location shows that curvature, shape and \ac{AUC} coincide and are almost identical.
On average, the \ac{FWHM} is $12$\% smaller for negative boli in the right atrium and $5$\% smaller in the left ventricle, which predicts shorter rise times for negative boli.
Boli were applied with identical procedure and alternatingly, hence a systematical fault in the setup is unlikely. An excessive low-pass filter setting would result in underestimating the bolus peak and equalize the \ac{FWHM} difference between bolus types.

\begin{figure}[t!]
    \centering
    \includegraphics[width=0.85\linewidth]{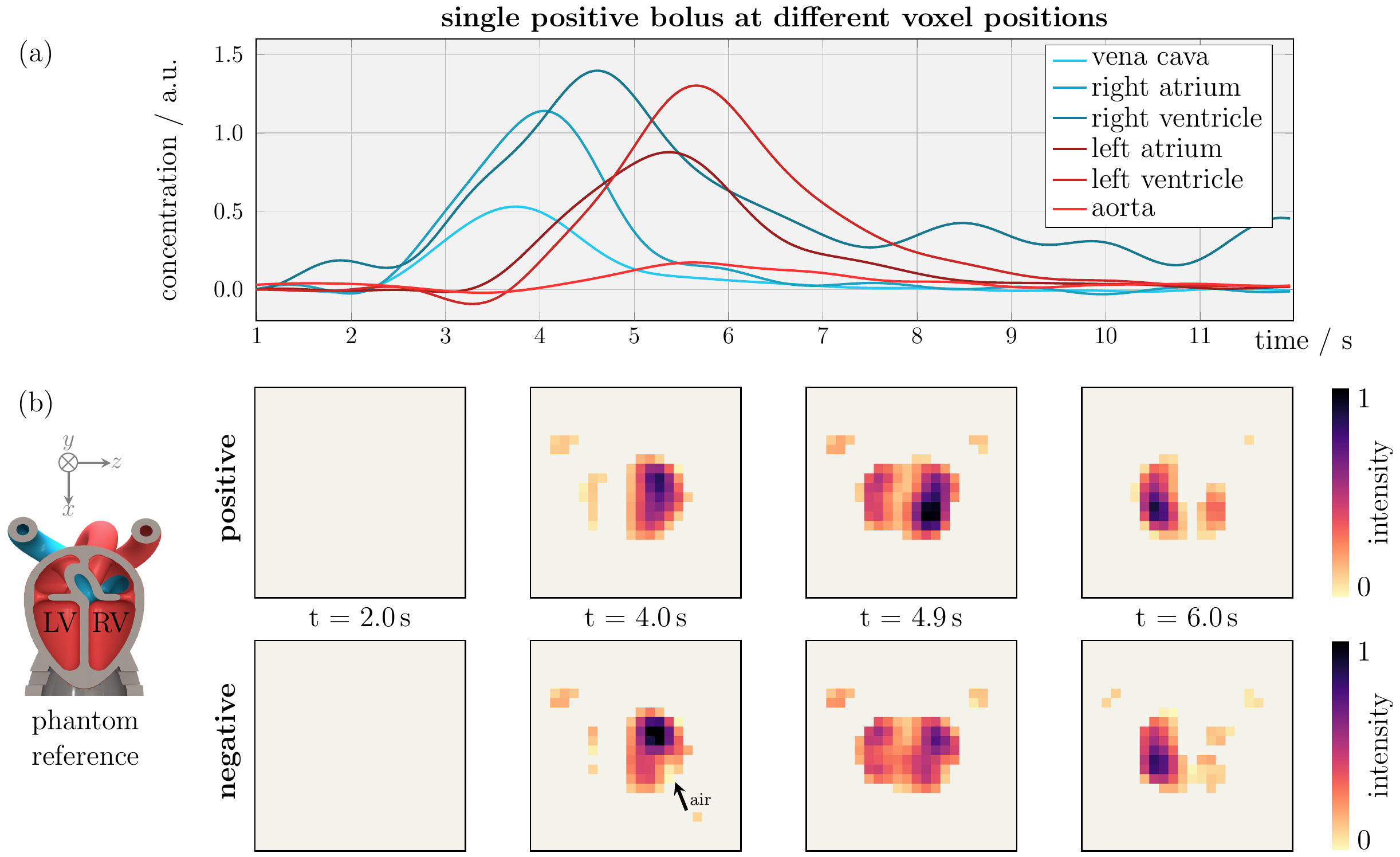} 
    \caption{\textbf{Local time responses and subtraction images.} In (a), the time response of a single bolus is tracked at different voxel locations in the heart phantom (filtered data). This is mapped in (b) below, where subtraction images are displayed for a positive and a negative bolus. A phantom reference slice on the left shows the orientation and layout of the 4 phantom heart chambers.}
    \label{fig:timeResp_subs}
\end{figure}

Dynamic time responses at different positions in the heart phantom are displayed in \autoref{fig:timeResp_subs}\,(a).
The delay, in other words the transition times between different areas, and the shape of the bolus peak can be distinguished. 
We expected the width of the peak to increase with time, the \ac{AUC} to remain constant, and the peak to be highest in the vena cava at the beginning. 
Limited resolution and unfortunate voxel spacing resulted in partial volume effects that led to locally lower concentrations compared with these expectations. 
The voxelsize of \qtyproduct{2x2x1}{mm^3} is the lower limit for the represented resolution, but the smallest features of the 3D printed phantom were tubes with diameters below \SI{2}{\mm}, e.g. for the vena cava, aorta and pulmonary vessels. As a result, these concentrations are underestimated for both bolus types, which is visible here in the maxima and \ac{AUC}s of the vena cava and aorta in \autoref{fig:timeResp_subs}\,(a).

Below in \autoref{fig:timeResp_subs}\,(b), subtraction images are shown, suggesting that negative boli do not show significant deviation from the positive ground truth and dynamic imaging is possible. Each bolus can be seen to enter the right ventricle first, then passing through the pulmonary circuit and on to the left side of the heart. 
A small air bubble accumulated over the course of the experiments in the right ventricle in spite of all efforts to prevent air from entering the phantom.
This caused a void in the images as no concentration changes occur except by movement of the bubble. Therefore, the bubble appears with zero intensity in subtraction images.

\begin{figure}[t!]
    \centering
    \includegraphics[width=1.0\linewidth]{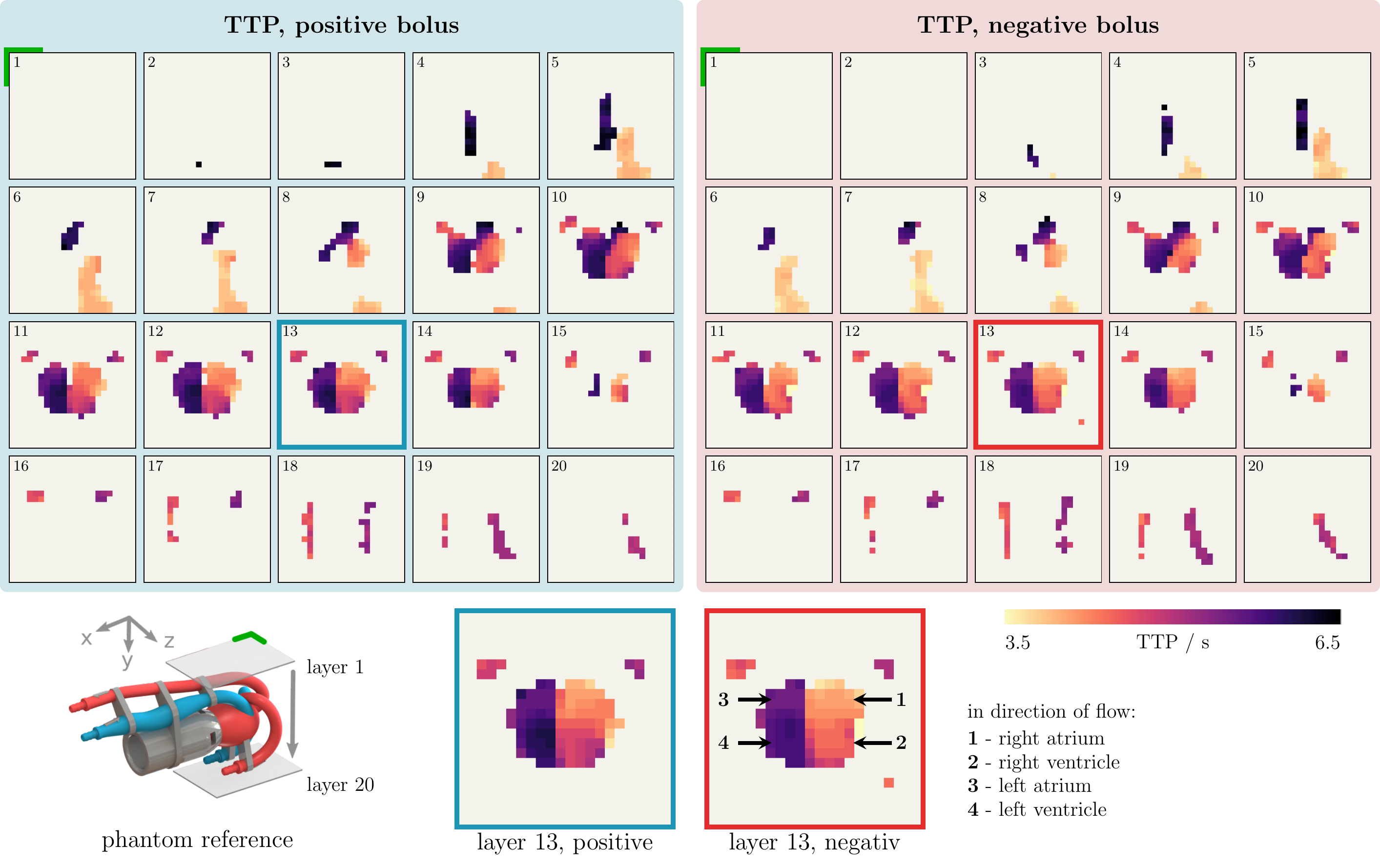}
    \caption{\textbf{Spatial progression of the \ac{TTP}.} On the left, a positive bolus was administered and on the right a negative bolus. The data were processed according to \autoref{fig:methods}. The time is mapped from bright colors ($3.5$\,s) to dark colors ($6.5$\,s) and 0\,s corresponds to the time of injection, identical to all time graphs above. The equidistant progression ($1$ to $20$) along the $y$-axis shows different $xz$-slices of the \ac{FOV}. Layer $13$ is magnified on the bottom and the $4$ heart chambers can be identified for each bolus type. The phantom reference slice is marked with a green corner for image orientation.}
    \label{fig:TTP}
\end{figure}

\subsection{Comparison of perfusion parameter maps of positive and negative boli}
\label{sec:res_maps}
In \autoref{fig:TTP}, the \ac{TTP} map is shown for a positive and a negative bolus, on the left and right, respectively. The top row shows the vena cava (blue) and the aorta (red), which are at the extremes of the colorbar, as they mark inlet and outlet of the heart phantom (compare to reference on bottom left). The bottom row shows the tubes of the pulmonary circuit (the return bend is outside the \ac{FOV}), which are very similar in color shape due to the small time difference (\ac{PTT}\,$=1$\,s). 
The center of the heart phantom (layer $13$) is enhanced to show that the $4$ chambers of the heart are identifiable by slight differences in shading. 
The \ac{TTP} was successfully obtained from a negative bolus and deviations between the two bolus types are low and in the same range of variations between administered boli of identical type.
Note that the filtered data used in these results is not normalized, as this would impact the \ac{rBF}, \ac{rBV}, \ac{MTT}, and the shape of the curve is therefore preserved.

\begin{figure}[t!]
    \centering
    \includegraphics[width=0.75\linewidth]{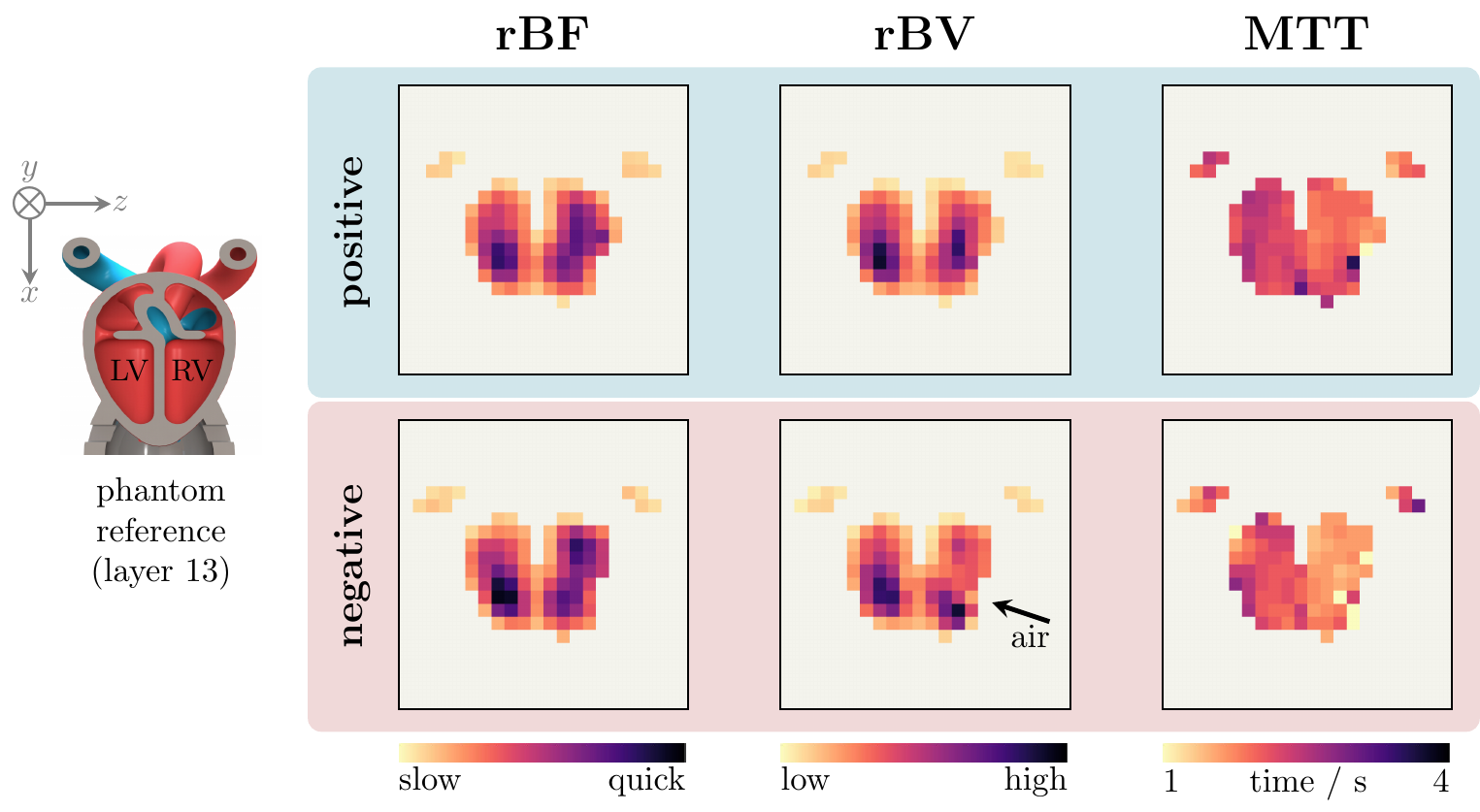}
    \caption{\textbf{The rBF, rBV and MTT for both bolus types.} Shown is a single slice (layer 13) for a positive and a negative bolus on the top and bottom respectively. Note that the images of each perfusion parameter are scaled to a single common colormap vertically and not normalized individually.}
    \label{fig:BFBVMTT}
\end{figure}

Further perfusion parameters, namely the \ac{rBF}, \ac{rBV} and \ac{MTT}, are shown in \autoref{fig:BFBVMTT}. Differences between the results of positive and negative bolus are small. The slightly darker tint of the negative \ac{rBF} corresponds to the above mentioned difference in the \ac{FWHM}, also visible in the data in \autoref{fig:timeResp_raw}\,(b). The same tendency can be seen in the \ac{MTT}, which is a little shorter for negative boli. However, the \ac{MTT} is generally around \SI{2.5}{\s} for both boli types and due to minor dispersion of the boli, the shading is darker in the left ventricle compared to the right ventricle, indicating that the bolus passed first through the right heart chamber and then through the left chamber. 
In the area of the pulmonary artery and the pulmonary vein, the actual blood flow is expected to be higher due to the smaller cross-section. The calculated blood flow shows the opposite due to the lower measured concentration. As already mentioned above with regard to \autoref{fig:timeResp_subs}\,(a), the local underestimation can be explained by partial volume effects and that the system matrix resolution is lower than the smallest vessels cross-section. This is consequently not a problem for the comparison between positive and negative boli, but a resolution-related limitation by the imaging system.
Low resolution and partial volume effects prevented a correct definition of the arterial input function, consequently the BF and BV were calculated in a relative manner. Note that \ac{rBF} and \ac{rBV} are relative, however, they are scaled identically for positive and negative bolus types. These two images for one perfusion parameter are not normalized individually, therefore they refer to a common colormap for direct comparability. 

As mentioned in \autoref{sec:phanmeas}, the tiny air pockets used to avoid dispersion during bolus preparation accumulated in the phantom, due to the position of the catheter just in front on the phantom. However, a bubble trap behind the catheter would have caused high dispersion of the bolus. Air also arose from cavitation in proximity to the pump, which prevented long experiments with more than $10$ boli (around \SI{5}{\min} each).

%%%%%%%%%%%%%%%%%%%%%%%%%%%%%%%%%%%%%%%%%%%%%%%%%%%%%%%%%%%%%%%%%%%%%%%%%
%%%%%%%%%%%%%%%%%%%%%%%%%%%%%%%%%%%%%%%%%%%%%%%%%%%%%%%%%%%%%%%%%%%%%%%%%
\section{Discussion}

We presented a novel method without the need of additional tracer dosage for calculating perfusion maps in \ac{MPI}. 
The method was evaluated based on phantom measurements, including the visibility of negative boli in the raw data (\autoref{fig:timeResp_raw}\,(a)), in normalized data (\autoref{fig:timeResp_raw}\,(b)), in dynamic subtraction images (\autoref{fig:timeResp_subs}\,(b)), and in perfusion maps (\autoref{fig:TTP} and \autoref{fig:BFBVMTT}). Overall deviations between the reconstructed images of a negative bolus and a positive bolus are low and in a similar range to the variations seen with multiple administrations of a positive bolus.

Regarding signal quality, negative boli do not need to be identical in volume to positive boli for perfusion imaging, they could be chosen larger for a better \ac{SNR}, since the human body is compatible with large amounts of neutral saline solution as opposed to magnetic particles.
The maximum achievable \ac{SNR} of a negative bolus depends on the baseline concentration in advance to the injection and on the injection volume of the bolus itself. If chosen in such a way that the negative bolus completely displaces all tracer from a \ac{VOI} during passing, maximum contrast is created. However, this is not desirable due to the intensity plateau that it creates, just as input-clipping of a large positive bolus should be avoided for the correct calculation of all perfusion parameters. A mindful selection of bolus volume and baseline concentration is necessary, similar to the selection of a sufficient dose for positive boli that is just visible for the planned imaging scenario. Further tracer-specific investigations regarding the baseline concentration for sufficient \textit{in-vivo} \ac{SNR} should be conducted.

Limitations to the efficiency and impact of the presented method are rooted in the availability and performance of a long circulating tracer for humans that upholds the mandatory baseline concentration. In addition, the \textit{in-vivo} performance of saline solution as a bolus has not been investigated. Some modifications to reduce dispersion in real blood may be required before the bolus reaches the desired organ (e.g. the brain). 

Although our study chose an exemplary scenario of a rat-heart phantom, and therefore a focus on cardiovascular imaging, the proof of concept to use negative boli for perfusion maps is transferable to other areas e.g. cerebrovascular imaging.
In a scenario that encompasses a long-term observation e.g. the monitoring of a future stroke patient, positive boli should be dosed as before, with a minimum of iron added for each acquisition of the perfusion parameters. Following several administered positive boli of a long circulating tracer, when a baseline concentration is reached, the observation can be continued using negative boli without adding to the particle iron mass in the body. A combination of positive-negative boli allows for a much longer observation with shorter intermediate boli time compared to positive boli only. Positive boli should be used if the blood pool concentration falls short of the baseline concentration for a negative bolus, caused by tracer which is taken up by the liver.
Another potential application is in endovascular interventions, e.g. stenosis assessment and stent implantation. The morphology of the vessel would be visible from the long circulating tracer and the catheter via a marker~\autocite{herz_magnetic_2019}. A small negative bolus could be administered directly in the field of view, without dispersion, to evaluate the procedure e.g. by calculating the blood flow. This can be repeated without adding more iron compared to positive bolus evaluation.

Generally valid quantitative statements about the amount of tracer reduction cannot be made on the basis of our experiments. The specific iron reduction by using negative boli depends on the surveillance scenario, the number of boli administered, the half-life of the tracer in blood, and other monitoring specific parameters.

In conclusion, our study proposes a method for obtaining perfusion images in \acl{MPI} with a positive-negative contrast sequence. The results from phantom experiments demonstrate that the negative bolus approach effectively contributes an image contrast resulting in concurrent signals with positive boli. Furthermore, our proposed method enables an increased monitoring time of a future patient while keeping the total iron dose constant. We believe that our results can significantly contribute to making MPI perfusion imaging techniques applicable in a clinical scenario with long monitoring times.

\section*{Acknowledgements}
The authors would like to thank Marija Boberg for proof reading and fruitful additions to the figures.
The authors thankfully acknowledge the financial support by the German Research Foundation (DFG, grant number KN 1108/7-1 and GR 5287/2-1).

%%%%%%%%%%%%%%%%%%%%%%%%%%%%%%%%%%%%%%%%%%%%%%%%%%%%%%%%%%%%%%%%%%%%%%%%%

\printbibliography

\vfill

\end{document}